\documentclass[fleqn,usenatbib]{mnras}
\usepackage{txfonts}
\usepackage{pinlabel}

\usepackage[T1]{fontenc}

\DeclareRobustCommand{\VAN}[3]{#2}
\let\VANthebibliography\thebibliography
\def\thebibliography{\DeclareRobustCommand{\VAN}[3]{##3}\VANthebibliography}

\title[Neutral-Neutral Synthesis in Cometary Comae]{Neutral-Neutral Synthesis of Organic Molecules in Cometary Comae}
\author[Cordiner \& Charnley]{
M. A. Cordiner,$^{1,2}$\thanks{E-mail: martin.cordiner@nasa.gov}
and S. B. Charnley$^{1}$
\\
$^1$Astrochemistry Laboratory, NASA Goddard Space Flight Center, 8800 Greenbelt Road, Greenbelt, MD 20771, USA.\\
$^2$Department of Physics, Catholic University of America, Washington, DC 20064, USA.
}

\date{Accepted for publication in MNRAS on April 16th 2021}

\pubyear{2021}

\begin{document}
\label{firstpage}
\pagerange{\pageref{firstpage}--\pageref{lastpage}}
\maketitle

%% Mark off the abstract in the ``abstract'' environment. 
\begin{abstract}

Remote and in-situ observations of cometary gases have revealed the presence of a wealth of complex organic molecules, including carbon chains, alcohols, imines and the amino acid glycine. Such chemical complexity in cometary material implies that impacts by comets could have supplied reagents for prebiotic chemistry to young planetary surfaces. However, the assumption that some of the molecules observed in cometary comae at millimetre wavelengths originate from ices stored inside the nucleus has not yet been proven. In fact, the comae of moderately-active comets reach sufficient densities within a few thousand kilometers of the nucleus for an active (solar radiation-driven) photochemistry to ensue. Here we present results from our latest chemical-hydrodynamic models incorporating an updated reaction network, and show that the commonly-observed HC$_3$N (cyanoacetylene) and NH$_2$CHO (formamide) molecules can be efficiently produced in cometary comae as a result of two-body, neutral-neutral, gas-phase reactions involving well-known coma gases. In the presence of a near-nucleus distributed source of CN (similar to that observed by the Rosetta spacecraft at comet 67P), we find that sufficient HC$_3$N and NH$_2$CHO can be synthesized to match the abundances of these molecules in previous observations of Oort Cloud comets. The precise origin of these (and other) complex organic molecules in cometary comae can be verified through interferometric mapping observations, for example, using the Atacama Large Millimeter/submillimeter Array (ALMA).

\end{abstract}


\section{Introduction}

Cometary ices are some of the most pristine, ancient materials in our Solar System. They have remained largely unaltered since they accreted during the birth of the planets (or earlier, in the interstellar medium), so their study provides unique information on the chemical conditions prevalent during the earliest history of our Solar System \citep{mum11,alt17}. Comets are also believed to have delivered volatiles and organics during impacts with planets, so understanding their compositions provides insight into the chemical regents that may have been present to drive prebiotic chemistry on the surfaces of Solar System bodies \citep{ehr00,mar10}.

Cometary ice abundances are derived primarily through multi-wavelength remote observations of their atmospheres/comae \citep{coc15}, while in-situ spacecraft missions provide further details on select comets (\emph{e.g.} \citealt{wyc88}; \citealt{rub19}). Coma mapping observations reveal that several well-known cometary molecules (including C$_2$, CN, HNC, H$_2$CO and others) have ``distributed'' sources, identified through their extended, relatively flat spatial profiles \citep{ahe95,cot08,cor14}, compared with the strongly centrally-peaked spatial profiles of species released directly from the nucleus. Interpreted within the \citet{has57} paradigm of an isotropically-expanding, constant-velocity outflow, it is clear that some ubiquitously-observed cometary molecules do not originate primarily from the nucleus, but instead arise in the coma, at distances of thousands to tens-of-thousands of kilometers from the nucleus. Coma ``daughter'' species  are believed to be produced from the breakdown of larger ``parent'' molecules (sublimated directly from the nucleus), as a result of photodissociation.  A plausible source for some of the observed distributed coma molecules is from the destruction of dust grains or large organic molecules in the coma \citep{cot08}. Recently, the amino acid glycine was detected by the Rosetta spacecraft during its mission to comet 67P Churyumov-Gerasimenko, with a radial density profile consistent with a distributed source, from the sublimation of ice-coated dust grains \citep{alt16,had19}.

Millimetre-wave spectroscopy has emerged as the leading ground-based method for identifying new coma molecules \citep[\emph{e.g.}][]{boc00,biv14}, including the first detections of cyanoacetylene (HC$_3$N), formamide (NH$_2$CHO), formic acid (HCOOH), ethylene glycol (C$_2$H$_6$O$_2$), and most recently: glycolaldehyde (C$_2$H$_4$O$_2$) and ethanol (C$_2$H$_5$OH) \citep{biv15}.  It is commonly assumed that they originate from the sublimation of ices stored inside the nucleus, but the presence of these species inside cometary nuclei is largely unproven. The majority of ground-based observations of organic molecules are made using single-pointing observations with a single-dish millimetre-wave telescope (such as the IRAM 30~m, with its $\sim10''$-diameter angular resolution; corresponding to $\sim7000$~km at 1~au Geocentric distance), and are therefore lacking in spatial information required to constrain the nucleus \emph{vs.} coma origins of these gases. 

\citet{rod01} used a chemical/hydrodynamic model (upon which our present study is based), to investigate the possible coma synthesis of four organic molecules (HCOOH, HCOOCH$_3$, HC$_3$N and CH$_3$CN), which were detected in comet C/1995 O1 (Hale-Bopp) using radio spectroscopy. They determined that coma chemistry was insufficient to reproduce the observed abundances of these species.   However, our latest models --- which now include distributed sources of CN and H$_2$CO in the coma --- show that some molecules can, in fact, be produced rapidly as a result of gas-phase chemical reactions. Here we present chemical model calculations for HC$_3$N and NH$_2$CHO, demonstrating the synthesis of these species in detectable quantities.

\section{Model}
\label{sec:model}

The details of our coma model were described by \citet{rod02}, and the chemical network has been updated for carbon, nitrogen and oxygen-bearing species, as well as including negative ion chemistry (see \citealt{cor14b}). The number densities ($n_i$) of 283 gas-phase species ($i$), and the temperatures ($T_x$) of four fluids (neutrals, electrons and positive and negative ions) are calculated as a function distance ($r$) from the nucleus, using the DVODE package \citep{hin19} to solve the following set of differential equations:

\begin{equation}
\frac{dn_i}{dr}=\frac{K_i}{v}-\frac{n_i}{v}\frac{dv}{dr}-\frac{2n_i}{r}
\end{equation}

\begin{equation}
\frac{dT_x}{dr}=\frac{(\gamma_x-1)T_x}{v}\left(\frac{G_x}{n_xk_BT_x}-\frac{2v}{r}-\frac{dv}{dr}-\frac{N_x}{(\gamma_x-1)n_x}\right)
\end{equation}

\begin{equation}
\frac{dv}{dr}=\frac{1}{\rho_s{v}^2-(\gamma{n}k_BT)_s}\left(F_sv-((\gamma-1)G)_s-M_sv^2+\frac{2v}{r}(\gamma{n}k_BT)_s\right)
\end{equation}

The chemical source terms ($K_i$) are equal to the sum over all production and loss rates for each species, $v$ is the coma outflow velocity, $T_x$ and $N_x$ are the respective temperature and density source terms for each fluid ($x$), and $\gamma_x$ are the ratios of specific heat (Laplace's coefficient) of the fluids. $G_x$ are the thermal energy source terms, which include contributions due to the the chemical reaction enthalpies, collisions between ions, neutrals and electrons, radiative energy loss from H$_2$O, and loss of fast H and H$_2$ from the coma. $M$ and $F$ are the mass and momentum source terms, respectively, and $\rho$ is the mass density; subscript $s$ implies summation over all four fluids in the model.

Parent gases are released from the nucleus and undergo isotropic expansion into the vacuum, reaching  $v=0.7$~km\,s$^{-1}$ by $r=1000$~km. The initial gas kinetic temperature was set to 100~K, with a nucleus radius of 2.5~km. The computed coma temperature and outflow velocity evolve in a very similar way to Figure 1 of \citet{rod02}.  Photodissociation of outflowing gases occurs due to solar radiation (using a nominal heliocentric distance $R_h=1$~au), and the ensuing photochemistry is calculated following a network of 3842 reactions. Abundances of parent molecules are chosen for a typical ``organic rich'' comet (see Table \ref{tab:parents}); based on observations at infrared \citep{del16} and radio \citep{boc17} wavelengths, and supplemented by in-situ measurements of O$_2$ and N$_2$ in comet 67P by \citet{bie15} and \citet{rub19}, respectively. We adopt an H$_2$O production rate of $5\times10^{29}$~s$^{-1}$ (released directly from the nucleus), which is representative of a moderately active comet at $R_h=1$~au. 

Our model has the ability to include distributed sources of molecules from the photolysis of (unidentified) parent species. In particular, we include distributed sources of H$_2$CO and CN (named $P1$ and $P2$, respectively), with photolysis rates of $\Gamma_{P1}=1.3\times10^{-4}$~s$^{-1}$ and $\Gamma_{P2}=3.23\times10^{-5}$~s$^{-1}$. These rates were chosen to be consistent with previous observations showing spatially-extended distributions for these gases, with Haser parent scale lengths (at $R_h=1$~au) of $L_p=5000$~km for H$_2$CO \citep{biv21} and $L_p=3.1\times10^4$~km for CN \citep{fra05}. An additional, inner-coma CN source ($P3$) is also included in some of our models  (see Section \ref{sec:results}), based on a recent analysis of CN data from comet 67P \citep{han20}.

The synthesis of cyanoacetylene (HC$_3$N) occurs in the coma \emph{via} the following neutral-neutral, gas-phase reaction, measured in the laboratory by \citet{lic86} and \citet{sim93}:

\begin{equation}
	{\rm CN} + {\rm C_2H_2} \longrightarrow {\rm HC_3N} + {\rm H}   
\end{equation}

Formamide (NH$_2$CHO) is produced by the neutral-neutral reaction:

\begin{equation}
	{\rm NH_2} + {\rm H_2CO} \longrightarrow {\rm NH_2CHO} + {\rm H}
\end{equation}

This reaction was studied theoretically by \citealt{bar15} and \citealt{sko17}, who calculated it to be efficient at the low (10--100~K) temperatures found in interstellar clouds and cometary comae; we adopt the lower estimate for the rate coefficient ($k=7.79\times10^{-15}(T/300)^{-2.56}e^{-4.9/T}$ cm$^3$\,s$^{-1}$), from the latter study.

\begin{table}
\centering
\caption{Parent Abundances with Respect to H$_2$O \label{tab:parents}}
\begin{tabular}{llll}
\hline
\hline
Species&Abundance&Species&Abundance\\
\hline
H$_2$O & 1.00    & CH$_4$ & 0.01 \\
CO$_2$ & 0.10    &  NH$_3$  & 0.01\\
CO     & 0.05    & H$_2$CO & $5.0\times10^{-3}$\\
O$_2$  & 0.04    & C$_2$H$_2$ & $3.0\times10^{-3}$\\
CH$_3$OH& 0.04   & HCN & $2.0\times10^{-3}$\\
C$_2$H$_6$ & 0.01& N$_2$ & $9.0\times10^{-4}$\\
\hline
\end{tabular}
\end{table}

\section{Results}

\subsection{HC$_3$N and NH$_2$CHO}
\label{sec:results}

\begin{figure*}
\labellist
\pinlabel a) at 6 192
\endlabellist
\includegraphics[width=0.49\textwidth]{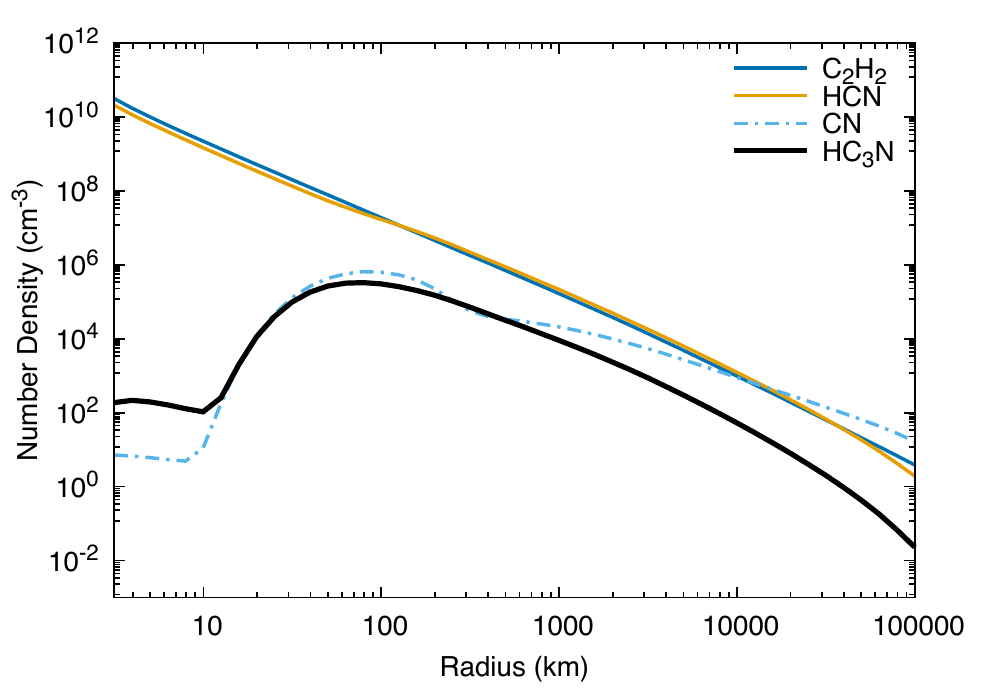}
\labellist
\pinlabel b) at 6 192
\endlabellist
\includegraphics[width=0.49\textwidth]{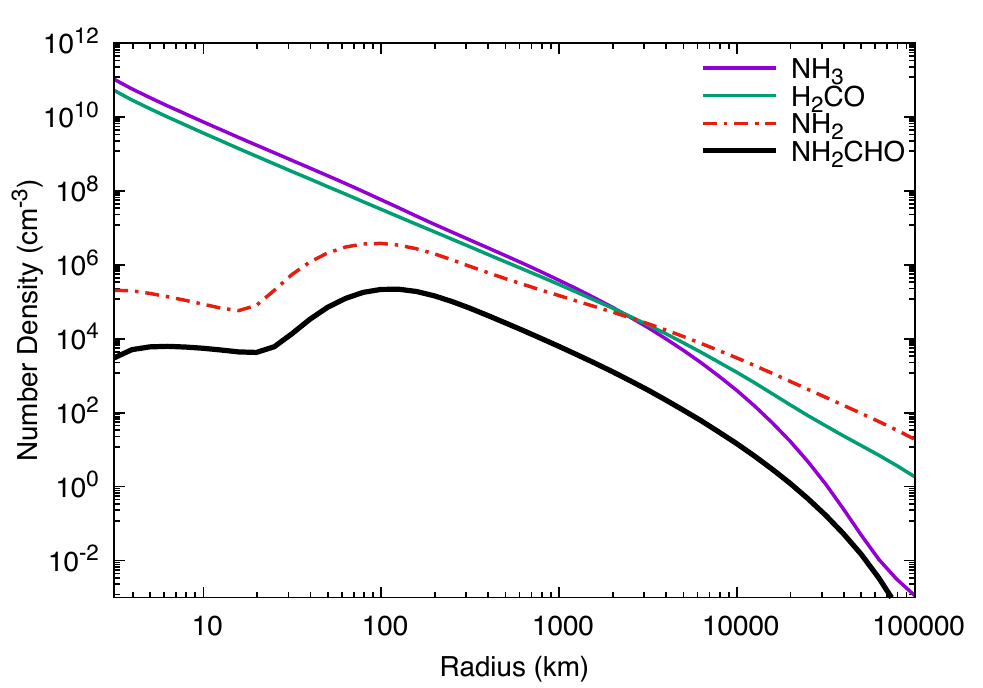}\\
\labellist
\pinlabel c) at 6 192
\endlabellist
\includegraphics[width=0.49\textwidth]{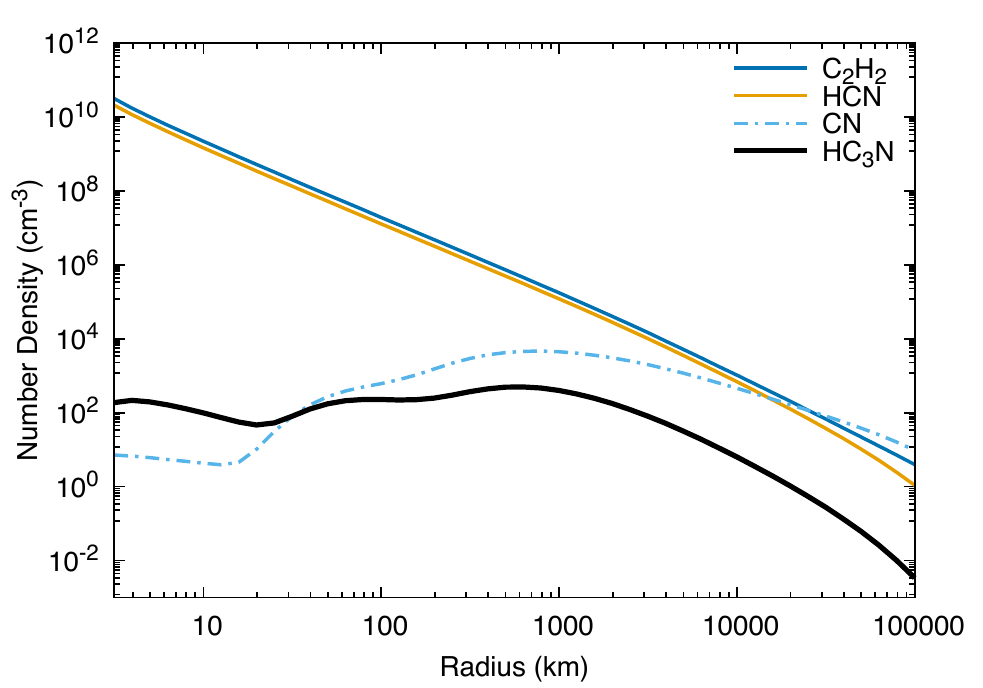}
\labellist
\pinlabel d) at 6 192
\endlabellist
\includegraphics[width=0.49\textwidth]{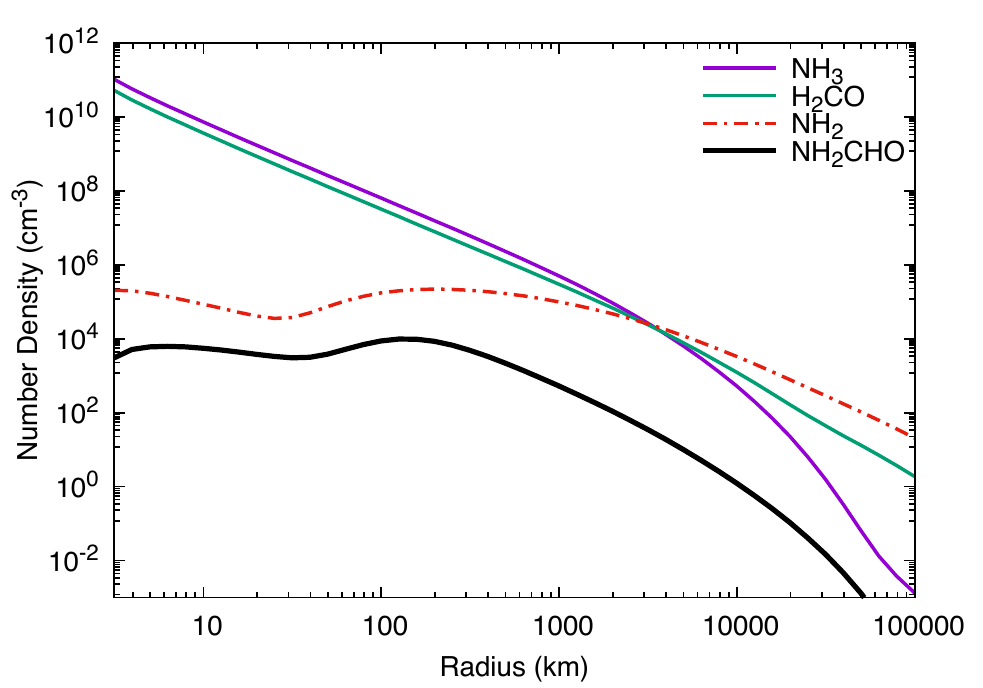}\\
\labellist
\pinlabel e) at 6 192
\endlabellist
\includegraphics[width=0.49\textwidth]{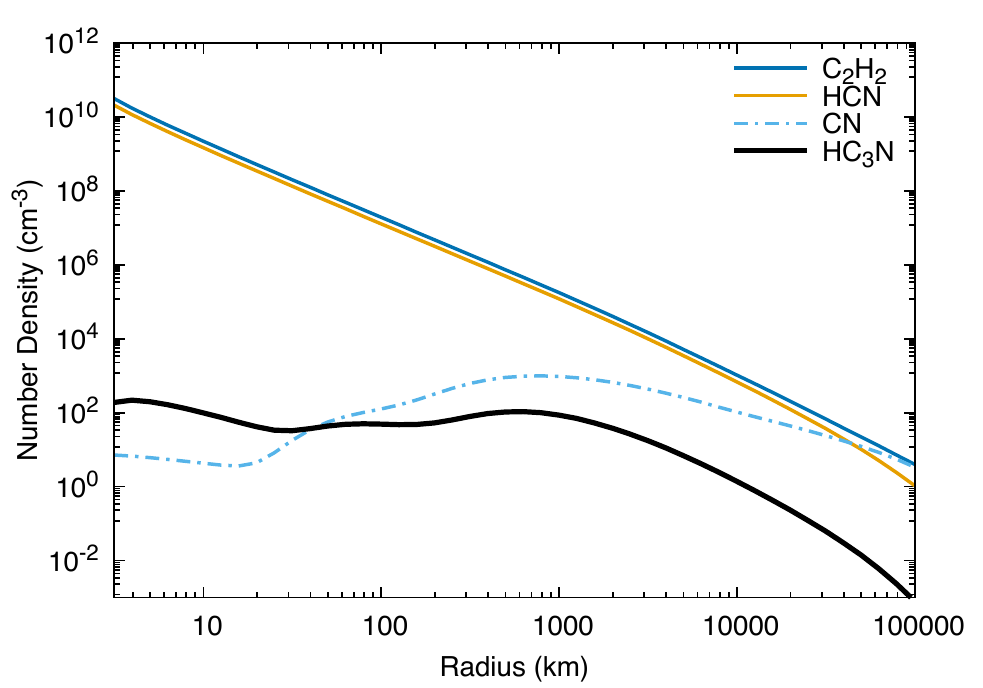}
\labellist
\pinlabel f) at 6 192
\endlabellist
\includegraphics[width=0.49\textwidth]{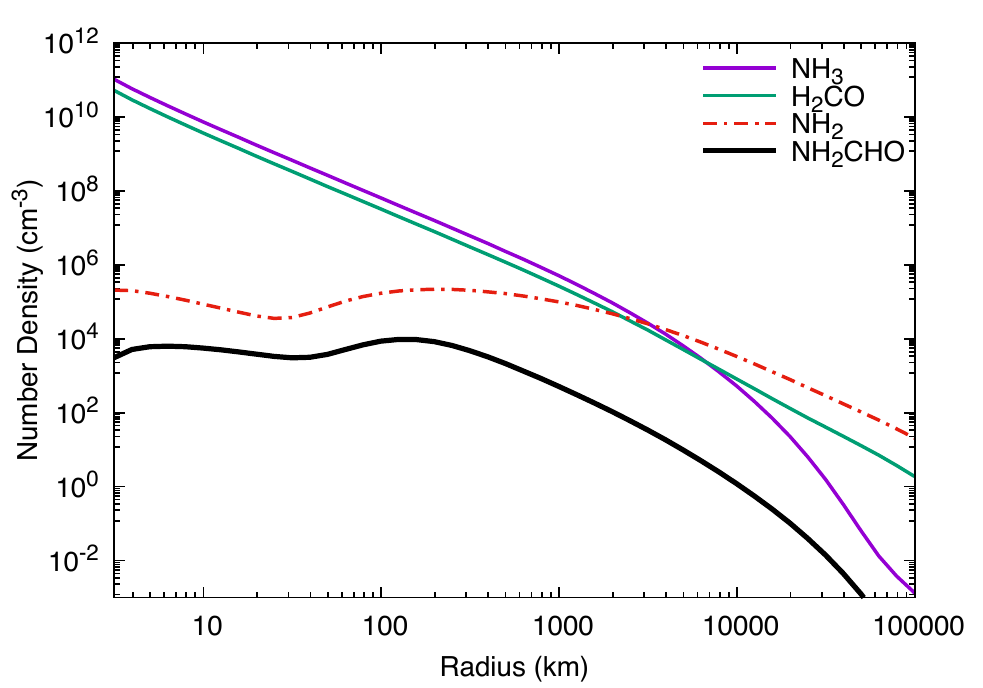}\\
\caption{Coma model output showing molecular number densities as a function of radius. Parent species are shown with solid, colored curves, and photolysis products are shown with dot-dashed line styles. Thick black curves show HC$_3$N (left column) and NH$_2$CHO (right column), formed as a result of Equations (4) and (5), respectively. Results from Model A (panels (a) and (b); top row) include a distributed source of H$_2$CO ($P1$) and two distributed sources of CN ($P2$ and $P3$). Model B are (panels (c) and (d); middle row), and have single distributed sources of CN ($P2$) and H$_2$CO ($P1$). Panels (e) and (f) (bottom row) are for Model C, which is the base model, with no additional sources of CN or H$_2$CO. \label{fig:models}}
\end{figure*}

Figure \ref{fig:models} shows the number densities ($n_i(r)$) of the molecules most relevant to this study, including parent species, photochemical daughters and products of coma chemistry. Three modeling scenarios are shown --- Model A (panels (a) and (b)) includes a distributed source of H$_2$CO from the breakdown of a large (unidentified) organic precursor molecule $P1$ at distances $r\sim5000$~km from the nucleus (see Section \ref{sec:model}), as well as two distributed sources of CN ($P2$ and $P3$), which give rise to peak gas-phase CN production at distances around $r\sim10^4$~km and $r\sim100$~km, respectively. Model B (panels (c) and (d)) has a single distributed source of H$_2$CO ($P1$) and a single, outer-coma distributed source of CN ($P2$). Model C (panels (e) and (f)) has no additional distributed sources of H$_2$CO or CN. In model C, H$_2$CO is a parent species (Table \ref{tab:parents}) with a relatively short photolysis scale-length of $\sim10^4$~km (due to its large photolysis rate $\Gamma_{\rm H_2CO}=2.3\times10^{-4}$~s$^{-1}$); however, there is still a significant contribution to the coma H$_2$CO abundance at distances $r\sim10^5$~km due to the photodissociation of methanol (CH$_3$OH + $h\nu$ $\longrightarrow$ H$_2$CO + H$_2$), which occurs at a slower rate \citep{hue15}. 

A significant source of the CN radical in all models is \emph{via} the photodissociation reaction HCN + $h\nu$ $\longrightarrow$ CN + H. Indeed, this is the dominant source of CN in Model C, but it was shown by \citet{boc85}, \citet{ahe95}, \citet{fra05} and \citet{del16} that the observed abundances of CN in comets are often inconsistent with a dominant origin from HCN photolysis. Furthermore, \citet{fra05} showed that the mean production scale length of CN in comets at $R_h<3$ au is not consistent with HCN photolysis, so an additional coma source of CN is indicated. Models A and B are therefore considered to better represent the chemistry of real comets. Using an initial abundance for the CN parent ($P2$) of 0.32 \% with respect to H$_2$O \citep{ahe95}, Model B reaches a peak CN density of $n_{\rm CN}=4.7\times10^3$~cm$^{-3}$ around 800~km from the nucleus, whereas Model C (with no additional CN parent) reaches only $n_{\rm CN}=1.0\times10^3$~cm$^{-3}$. The elevated CN density translates almost linearly to a larger HC$_3$N abundance \emph{via} Equation (4). The distributed H$_2$CO source (from the breakdown of parent species $P1$) has a less noticeable impact on the overall coma H$_2$CO number density, which reaches $n_{\rm H_2CO}=1.3\times10^3$~cm$^{-3}$ at $r=10^4$~km in Model B and $n_{\rm H_2CO}=0.8\times10^3$~cm$^{-3}$ in Model C.

The total yield of HC$_3$N and NH$_2$CHO in our models is quantified using an ``effective production rate'' $Q_e$. This quantity is closely related to the observationally-derived production rate ($Q$), and is defined as the production rate that would be measured from our model using a telescope beam diameter of $10''$ (the beam size of the IRAM 30-m telescope at 250~GHz), at a cometocentric distance of $\Delta=1$~au, under the assumption that the molecule of interest originates from the nucleus. The latter assumption has been employed in all previously published HC$_3$N and NH$_2$CHO production rates derived from ground-based mm-wave observations, so $Q_e$ values allow our model output to be compared directly with observations taken at a similar cometocentric distance ($\Delta\sim1$~au). The $Q_e$ values (Table \ref{tab:abunds}) are calculated by comparing our modeled column densities (averaged over a $10''$ beam) with a Haser parent model, using an outflow velocity equal to the column density-weighted mean HCN outflow velocity from our model ($\langle{v}\rangle=0.74$~km\,s$^{-1}$), and photolysis rates from \citet{hue15} and \citet{hea17}. The HCN molecule is used for this purpose since HCN is a common probe of the coma outflow velocity at mm/sub-mm wavelengths, but use of other parent molecules such as CH$_3$OH and H$_2$O to measure $\langle{v}\rangle$ produces very similar results. The $Q_e$ values are then converted to effective abundances ($a_e$) by taking the ratio with respect to the model H$_2$O production rate of $5\times10^{29}$ s$^{-1}$.

\begin{table*}
\caption{Effective Production Rates ($Q_e$) and Abundances ($a_e=Q_e/Q_{\rm H_2O}$) from our Coma Models\label{tab:abunds}}
\begin{tabular}{lcccccc}
\hline\hline
&\multicolumn{2}{c}{Model A} & \multicolumn{2}{c}{Model B} & \multicolumn{2}{c}{Model C}\\
\hline
Species & $Q_e$      & $a_e$   & $Q_e$      & $a_e$  & $Q_e$      & $a_e$   \\
        &($10^{24}$~s$^{-1}$) & (\%) & ($10^{24}$~s$^{-1}$) & (\%) & ($10^{24}$~s$^{-1}$) & (\%)\\
\hline
HC$_3$N  & 87.3          & $1.8\times10^{-2}$ & 6.7 & $1.4\times10^{-3}$ & 1.5 & $2.9\times10^{-4}$\\
NH$_2$CHO& 58.1          & $1.2\times10^{-2}$ & 4.9 &$1.0\times10^{-3}$  & 4.7 & $9.4\times10^{-4}$\\
\hline
\end{tabular}
\end{table*}

Upon inclusion of a distributed source of CN ($P2$), the effective HC$_3$N production rate increases by a factor of 5 from $1.5\times10^{24}$~s$^{-1}$ to $6.7\times10^{24}$~s$^{-1}$, whereas a distributed H$_2$CO source ($P1$) only results in a modest, 4\% increase to the NH$_2$CHO production rate. The corresponding effective abundances for model B (see Table \ref{tab:abunds}) of $1.4\times10^{-3}$\,\% for HC$_3$N and $1.0\times10^{-3}$\,\% for NH$_2$CHO are significant --- indeed, potentially detectable --- but are still less than the lowest values reported in the literature for these molecules in Oort Cloud comets ($2\times10^{-3}$\,\% and $8\times10^{-3}$\,\%, respectively; \citealt{boc17}). Oort Cloud comets (OCCs) have relatively long orbital periods, covering distances up to hundreds of thousands of astronomical units from the Sun. They are observationally distinct from the Jupiter Family Comets (JFCs), which orbit at distances $R_h\lesssim10$~au and are exposed to more frequent, repeated heating during successive perihelion passages. Statistical studies show that JFCs tend to be depleted relative to OCCs in some volatiles such as CH$_4$, C$_2$H$_2$ and CO \citep{del16}, and before the Rosetta mission to JFC 67P, there were no reported detections of HC$_3$N or NH$_2$CHO in any JFC. Both these molecules were detected early in the Rosetta mission by \citet{ler15} around $R_h=3.1$ au, but a more useful comparison is with the values measured by \citet{rub19} closer to perihelion ($R_h\sim1.5$ au) when the comet was more fully activated; these latter observations obtained HC$_3$N/H$_2$O = $4\times10^{-4}$\,\% and NH$_2$CHO/H$_2$O = $4\times10^{-3}$\,\%, respectively. Direct comparison between our model results and the measurements of comet 67P, however, requires the impact of the lower H$_2$O production rate of this comet to be considered (see Section \ref{sec:disc}).

A more prominent increase is seen in our model for both HC$_3$N and NH$_2$CHO when including a second distributed source of CN in the inner coma ($P3$). The properties of this additional source  were chosen to match in-situ measurements of comet 67P (Churyumov-Gerasimenko) by the Rosetta spacecraft \citep{han20}. Throughout the mission to 67P, the ROSINA mass spectrometer detected a clear signal due to gas-phase CN at cometocentric distances 20--200~km, with a local abundance ratio of $\sim0.01$--0.1 with respect to HCN at dates around perihelion. The CN spatial distribution was observed to be relatively flat compared with H$_2$O and other parent species in the coma, consistent with a distributed source of CN, with an abundance far in excess of what could be explained by photolysis of any known coma nitriles (see also \citealt{han21}). The observed excess CN signal was tentatively attributed by \citet{han20} to the breakdown of CN-bearing refractory particles, such as salts (ammonium cyanide --- NH$_4$CN, for example), nitrogen-rich dust, or organic macromolecules. The production rate ($Q_{P3}/Q_{\rm H_2O}=6\times10^{-3}$) and photolysis rate ($\Gamma_{P3}=3.5\times10^{-2}$~s$^{-1}$) of this additional (unidentified) CN parent were chosen in our model to produce a maximum $n_{\rm CN}/n_{\rm HCN}$ ratio of $\approx0.05$ between $r=100$--200~km from the nucleus, in line with the ROSINA measurements close to perihelion.

As a consequence of the increased CN density in the inner coma, Equation (4) proceeds more rapidly and the HC$_3$N effective abundance in Model A increases to 0.018\%. The presence of additional CN has several other knock-on effects for the coma chemistry. In particular, the CN radical quickly reacts with NH$_3$ (sublimating from the nucleus) to produce HCN + NH$_2$ \citep{sim94,tal09}. Some of this additional NH$_2$ reacts with H$_2$CO (\emph{via} Equation 5) to produce NH$_2$CHO, leading to a significant (factor of 12) increase in the NH$_2$CHO abundance, which then reaches $a_e=0.012$\%. Consequently, in Model A, both HC$_3$N and NH$_2$CHO attain abundances with respect to H$_2$O that are well within the range of values previously detected in Oort Cloud comets (0.002--0.068\% for HC$_3$N and 0.008--0.021\% for NH$_2$CHO; \citealt{boc17}), with no need for a nucleus (parent) source of either molecule.

\subsection{Other Molecules}

The reaction between CN and O$_2$ was studied in the laboratory by \citet{sim94}, and found to proceed rapidly at low temperatures. \citealt{fen09} determined OCN + O to be the dominant product channel, so large amounts of OCN are produced in our models. The effective OCN abundance inside a $10''$ beam is $a_e=0.2$\% in Model A and $a_e=4\times10^{-3}$\,\% in Model B, assuming a generic photolysis rate of $\Gamma_{\rm OCS}=10^{-5}$~s$^{-1}$. Gas-phase OCN was detected by the Rosetta spacecraft in the coma of comet 67P during a dust impact event \citep{alt20}, but has not yet been detected in any other comets. It has a moderate dipole moment of 0.64~D, with several rotational transitions in the millimetre and submillimetre range. Consequently, if the coma CN and O$_2$ abundances are similar to those measured in comet 67P close to perihelion, OCN may be bright enough to be detectable in a sufficiently active Oort Cloud comet. Indeed, since CN + O$_2$ is the main pathway to OCN in our model (other pathways are negligible), a measurement of the OCN production rate could be used to infer the product of CN $\times$ O$_2$ abundances in the inner coma of future comets, thus providing an indirect measurement of the O$_2$ abundance, which has so-far been impossible to obtain with ground-based observations. 

A smaller, but still significant, amount of NO is also produced as a result of the alternative product channel CN + O$_2$ $\longrightarrow$ CO + NO. The resulting NO goes on to react with OCN to produce N$_2$O and CO.

In addition to the main product channel (leading to NH$_2$ + HCN), the reaction between CN + NH$_3$ was suggested by \citet{her94} as a possible source of NH$_2$CN (cyanamide) in the interstellar medium. If this reaction proceeds with reasonable efficiency, NH$_2$CN could reach detectable abundances in the inner coma (0.08\% in Model A, assuming an NH$_2$CN product branching ratio of 1/3).  More recent quantum calculations by \citet{tal09}, however, indicate that this product channel may be negligible in comparison to the NH$_2$ + HCN channel.

\subsection{Synthetic Coma Maps}

\begin{figure}
\centering
\includegraphics[width=\columnwidth]{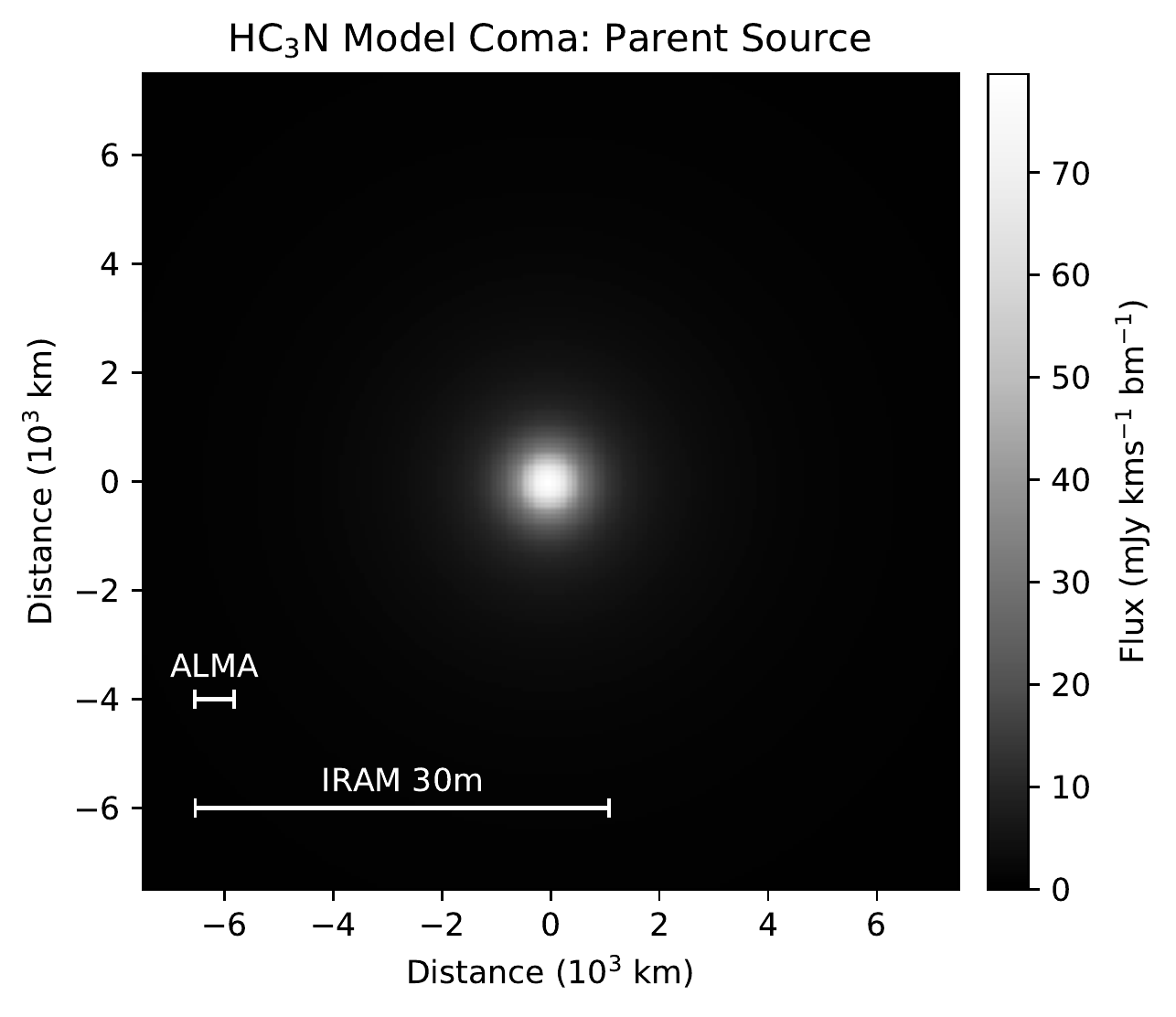}\\
\includegraphics[width=\columnwidth]{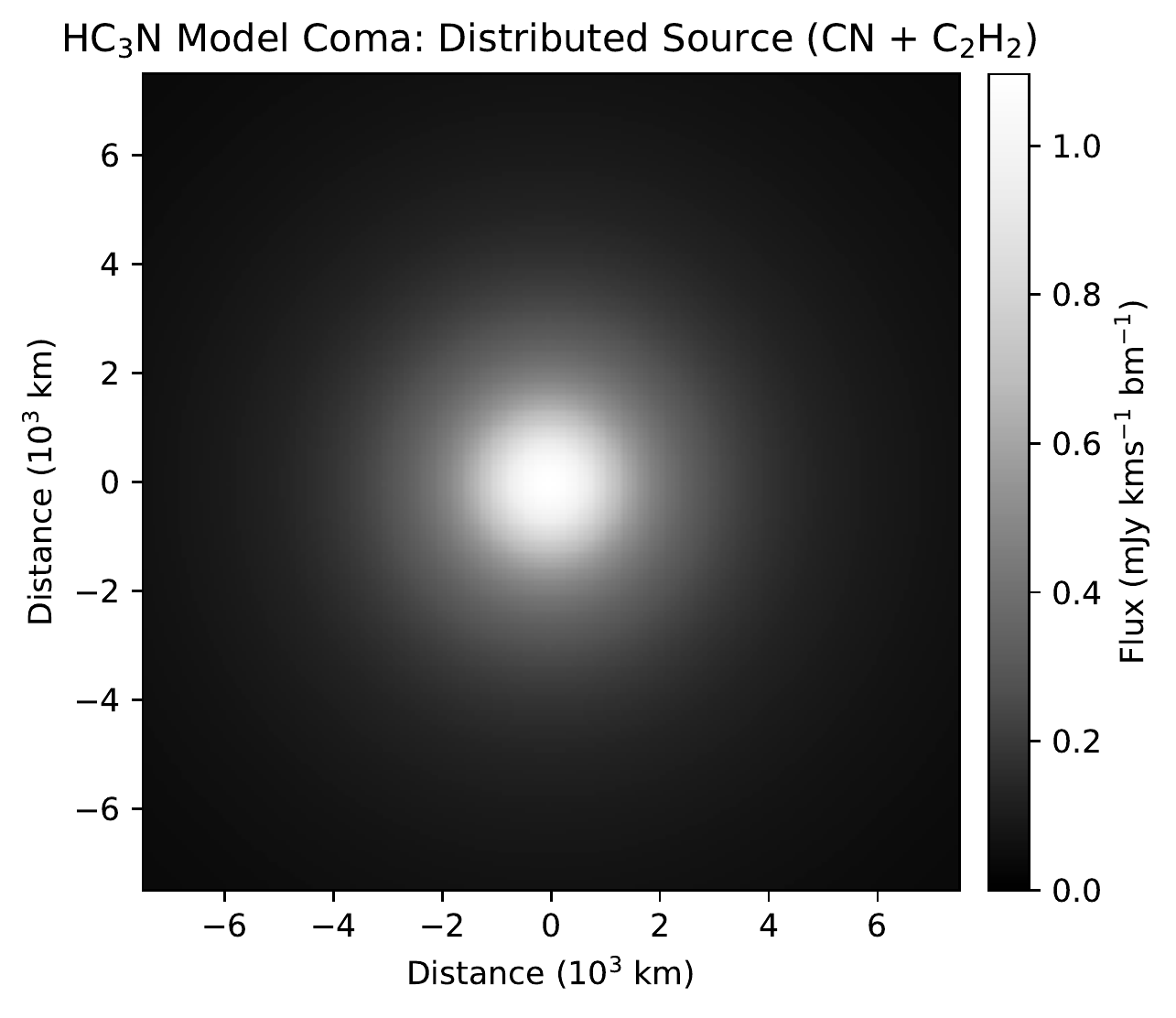}
\caption{Simulated HC$_3$N coma images (flux maps) for the $J=26-25$ transition at 236.5~GHz, assuming an excitation temperature of 50~K, smoothed to an angular resolution (beam FWHM) of $1''$. The top panel is for HC$_3$N as a parent (sublimating directly from the nucleus with $Q({\rm HC_3N})=10^{26}$~s$^{-1}$). Bottom panel is for HC$_3$N produced \emph{via} Equation (4), according to the model density profile in Figure \ref{fig:models} (panel c). Horizontal scale bars in the top panel indicate the characteristic spatial resolution of the ALMA and IRAM 30-m telescopes ($1''$ and $10''$, respectively). \label{fig:maps}}
\end{figure}

Our predictions regarding the origins of cometary HC$_3$N and NH$_2$CHO (and other molecules) in comets, may be tested through mapping observations of their spatial distributions with respect to the nucleus. Parent molecules (released directly from the nucleus) have compact brightness distributions that fall rapidly as $\sim1/\rho$, where $\rho$ is the nucleocentric distance projected in the plane of the sky. Daughter species (or products of coma chemistry), on the other hand, have flatter, more extended distributions. These two scenarios can be readily distinguished through interferometric millimetre/submillimetre observations (for example, using the ALMA telescope; \citealt{cor14}), or using a single-dish facility, in the case of an unusually close-up apparition (such as 252P/Linear in 2016, which reached a Geocentric distance of only 0.04~au; \citealt{cou17}). Detailed coma mapping observations can also provide crucial information on production scale-lengths for comparison with numerical models, to help elucidate the specific production pathway for the species in question. 

Synthetic mm-wave flux maps are shown for the case of HC$_3$N as a parent molecule in Figure \ref{fig:maps} (top panel), and for HC$_3$N produced from the reaction of CN with C$_2$H$_2$ according to Model B (bottom panel). These maps were constructed for the $J=26-25$ transition by raytracing the 3D projection of our model density output along the observer-comet sightline, using the RATRAN Sky code \citep{hog00}. An excitation temperature of 50~K and Geocentric distance $\Delta=1$~au were assumed, and the resulting maps were integrated along the spectral axis, then convolved with a Gaussian beam of FWHM~$=1''$, similar to ALMA's lowest resolution at this frequency. Distributed sources with parent scale lengths less than a few thousand kilometers are difficult to map reliably using a single-dish facility such as the IRAM 30 m telescope, as shown by the extent of the scale bars relative to the synthetic brightness distributions in Figure \ref{fig:maps}.  ALMA observations, on the other hand, are readily able to distinguish between the two scenarios, and can therefore provide proof of a nucleus or coma origin for HC$_3$N, NH$_2$CHO and other species.

Previous analyses of observational data obtained using radio interferometry of cometary gases have relied on simple Haser models to interpret the observed spatial profiles for distributed coma species such as H$_2$CO, HNC and CS \citep{boi07,cor14,cor17}. This approach assumes the detected molecules arise only as a result of photodissociation of a single parent species, and that the parent species (as well as the daughter species in question) does not have any additional sources or sinks as a result of chemical reactions or dust sources in the coma. Coma chemistry results in radial density profiles that can be distinctly different from Haser-model density profiles (see Figure \ref{fig:models}), so sufficiently detailed (high signal-to-noise) ALMA observations should have the ability to distinguish between different chemical production scenarios, as well as photodissociation, production from icy grains, and sublimation from the nucleus (or a combination these mechanisms).

\section{Discussion}
\label{sec:disc}

Our chemical models show that significant quantities of HC$_3$N, NH$_2$CHO and other molecules can be synthesized in the coma through neutral-neutral reactions involving simple chemical precursor species (CN, C$_2$H$_2$, NH$_2$, O$_2$ and H$_2$CO), already known to be abundant from previous comet observations. The presence of a distributed source of CN in the inner coma (as observed by \citealt{han20} for comet 67P), strongly enhances the efficiency of both reactions (4) and (5), bringing the effective abundances for Model A (see Table \ref{tab:abunds}) into the range of previously observed values for these molecules. The presence of such an additional CN source is consistent not only with Rosetta measurements of comet 67P, but also with previous studies regarding the (still unidentified) source of the closely-related HNC molecule in comets (see \citealt{cor17}, and references therein).

When HNC was first detected in comet C/1995 O1 (Hale-Bopp), the large HNC/HCN mixing ratio (similar to the value found in the interstellar medium) was taken as evidence that pristine (unprocessed) interstellar material may be incorporated into cometary nuclei. However, maps of the HNC spatial distribution in comet Hale-Bopp \citep{bla99}, as well as strong variability of the HNC/HCN ratio with heliocentric distance in a sample of 14 comets \citep{lis08}, pointed towards a distributed (coma) source of HNC, which was hypothesized to originate from the breakdown of macromolecular or dust precursor material \citep[see also][]{rod01b}. Detailed interferometric mapping using the Atacama Large Millimeter/submillimeter Array (ALMA) provided definitive evidence for a distributed source of HNC at distances of a few hundred kilometers from the nucleus of comet C/2013 S1 (ISON), leading to the conclusion that this molecule most likely originates from the degradation of nitrogen-rich organic refractory material \citep{cor14,cor17}.

Low-mass, refractory organic particles (CHON grains) were detected in large abundances in comet 1P/Halley by the Giotto mission \citep{kis86}, and \citet{wyc91} determined that 90\% of this comet's nitrogen budget was contained within the refractory dust component.  More recent in-situ work on comet 67P using the Rosetta COSIMA instrument revealed the presence of very large macromolecular compounds in the cometary dust, analogous to the insoluble organic matter present in carbonaceous meteorites \citep{fra16}.  With a mean N/C ratio of 3.5\% \citep{fra17}, this material presents a plausible source of additional CN radicals in the coma, among other small N-bearing organics. Macromolecules such as HCN polymer or hexamethylenetetramine (HMT, recently detected in meteorites by \citealt{oba20}) present another possible source of coma CN, but these specific compounds are yet to be found in cometary samples. Polyoxymethylene (POM, or formaldehyde polymer) has been studied extensively as another plausible macromolecular grain component in comets, and detailed models show that POM could explain the distributed H$_2$CO sources in comets 1P/Halley and C/1995 O1 (Hale–Bopp) \citep{cot04,fra06}.

\citet{han20} also considered the dissociation of NH$_4$CN as a plausible source of CN in the inner coma of comet 67P. The NH$_4$CN salt (also known as NH$_4$$^+$CN$^-$) is unstable in the gas phase, but could be carried into the coma as a solid embedded in (or adsorbed on the surface of) dust grains, before sublimating and dissociating. The possibility of ammoniated salts as a reservoir of HCN and NH$_3$ in comets was first discussed by \citet{mum18}. \citet{alt20} reported the likely presence of five different ammonium salts (NH$_4$$^+$X$^-$, where X$^-$ is a deprotonated acid, such as Cl$^-$, NCO$^-$ or HCOO$^-$) in the coma of comet 67P, based on ROSINA mass spectrometry during a dust impact event. Evidence for abundant ammonium salts on 67P's surface (including NH$_4$CN) was also found with Rosetta infrared spectroscopy \citep{poc20}.

While HCN + NH$_3$ is believed to be the dominant dissociation channel of NH$_4$CN, other products are possible \citep{alt20,han20}. By analogy with the dissociation of NH$_4$Cl, which has been studied using ab-initio methods, as well as in the laboratory (see \citealt{han19} and references therein), it is suggested that CN and CN$^-$ could be products of NH$_4$CN dissociation. Our models show that the majority of CN$^-$ produced in the inner coma from the spontaneous dissociation reaction NH$_4$CN $\longrightarrow$ NH$_4$$^+$ + CN$^-$ would be converted almost immediately to neutral CN (+ H) upon collision with NH$_4$$^+$ \citep[see][]{har08}. This therefore represents another possible source of CN to drive the neutral-neutral synthesis of HC$_3$N and NH$_2$CHO. More laboratory studies of NH$_4$CN dissociation products are needed to confirm this hypothesis.

In light of the recent detections of salts in comet 67P, the results of \citet{han19} and \citet{alt20} imply a possible contribution to the NH$_2$CHO coma abundance from NH$_4$COOH (ammonium formate) salt dissociation.  Our models show that the presence of additional sources of molecules in the inner coma (such as NH$_3$ and HCN), from the dissociation of salts including ammonium formate, ammonium cyanide and ammonium cyanate \citep[see][]{alt20}, lead to increased abundances of complex organics as a result of subsequent gas-phase reactions. Further investigations of the coma chemistry arising from such reactions are therefore warranted. The injection of ions, radicals and other reactive neutrals into the inner coma, following the dissociation of ammonium salts, would give rise to further, previously unstudied chemical processes in the outflowing cometary gas. To model this in detail, however, would require improved knowledge of the initial salt abundances and their full range of dissociation products, as well as the inclusion of a population of outflowing, sublimating grains into our model, which is beyond the scope of the present study.

%Additional uncertainties in the observed $Q$ values due to differing distances Delta.

Our findings for HC$_3$N are in contrast to \citet{rod01}, whose model predicted a low HC$_3$N/H$_2$O abundance ratio (averaged inside an $11''$ beam) of $7\times10^{-4}$\,\% for Hale-Bopp, compared with the observed value of $1.7\times10^{-2}$\,\%. In their model, HC$_3$N formation was driven by the CN + C$_2$H$_2$ reaction (4), with CN produced photolytically from HCN released directly from the nucleus. Consequently, they concluded that the observed HC$_3$N could not be synthesized in the coma by Equation (4) alone, and is therefore likely to be present in the nuclear ice. The \citet{rod01} model should have been more efficient than ours at producing HC$_3$N due to the omission of the (experimentally observed) 20~K activation energy barrier in Equation (4), allowing the reaction to proceed more rapidly at low temperatures in their model.  The ability of our new model to effectively reproduce the HC$_3$N observations is attributed primarily to the inclusion of the two distributed sources of CN ($P2$ and $P3$) in the coma for Model A, which were not considered by \citet{rod01}. 

It is interesting to note that HC$_3$N/H$_2$O ratio in comet 67P ($4\times10^{-4}$; \citealt{rub19}) is significantly lower than in the sample of Oort Cloud comets observed to-date ($2\times10^{-3}$–-$6.8\times10^{-4}$; \citealt{boc17}). Our modeling work provides a natural explanations for this, without requiring peculiar/anomalously-low HC$_3$N abundances in 67P's nucleus compared with OCCs. To investigate the production of HC$_3$N (and NH$_2$CHO) in lower-density coma environments such as that of 67P, we ran additional model calculations spanning a range of water production rates $Q_{\rm H_2O}=3\times10^{27}$--$5\times10^{30}$~s$^{-1}$ (covering the range of values observed in typical cometary apparitions). The chemical reaction rates for formation and destruction of the species of interest in our study vary as nonlinear functions of the coma density and temperature, so the changes in their coma abundances in response to $Q_{\rm H_2O}$ is nontrivial. The results of our tests show that for $Q_{\rm H_2O}$ above a few times $10^{29}$~s$^{-1}$, the HC$_3$N and NH$_2$CHO abundances with respect to water are not strongly dependent on the production rate (they vary by $<15$\% in this $Q_{\rm H_2O}$ range). For $Q_{\rm H_2O}\lesssim10^{29}$~s$^{-1}$, the size of the region dense enough for rapid neutral-neutral reactions to occur shrinks considerably, leading to significant reductions in the yields of HC$_3$N and NH$_2$CHO \emph{via} reactions (4) and (5).

To simulate comet 67P for comparison with Rosetta mass spectrometry measurements, we ran the $Q_{\rm H_2O}=3\times10^{27}$~s$^{-1}$ model at $R_h=1.5$~au, using \citet{rub19}'s measured parent abundances and observational circumstances (C$_2$H$_2$ was not measured by \citealt{rub19}, so we adopt the mean JFC mixing ratio of 0.07\% for this species, from \citealt{del16}). The inner-coma CN production rate was scaled to reproduce the CN/HCN ratio measured by \citet{han20} around the time of the \citet{rub19} measurements, and our modeled abundances ($a({\rm X})$, with respect to H$_2$O) were extracted at a distance of 220~km from the nucleus, which is close to the average distance at which the \citet{rub19} measurements were obtained. For Model A, we find $a({\rm HC_3N})=3.0\times10^{-4}$\,\% and $a({\rm NH_2CHO})=1.5\times10^{-5}$\,\%; for model B: $a({\rm HC_3N})=5.5\times10^{-6}$\,\% and $a({\rm NH_2CHO})=1.1\times10^{-5}$\,\%; and for model C: $a({\rm HC_3N})=8.9\times10^{-7}$\,\% and $a({\rm NH_2CHO})=1.1\times10^{-5}$\,\%. These values can be compared with the Rosetta measurements of $a({\rm HC_3N})=(4\pm2)\times10^{-4}$\,\% and $a({\rm NH_2CHO})=(4\pm2)\times10^{-3}$\,\%. The agreement for HC$_3$N (within the observational uncertainties) using Model A is surprisingly good considering the uncertain radial density profile of the inner-coma CN source. Our HC$_3$N/HCN ratio of $2.1\times10^{-3}$ is also in agreement with the value of $(2.9\pm1.9)\times10^{-3}$ from \citet{rub19}, but is somewhat less that the mean value of $(4.6\pm0.8)\times10^{-3}$ measured by \citet{han21} between $R_h=1.24$--1.74~au. This discrepancy could be accounted for by variability in the CN, HCN and C$_2$H$_2$ abundances during the extended (5 month) time period covered by the \citet{han21} study.  We note, however, that purely gas-phase chemistry would be expected to give rise to a steeper slope in the HC$_3$N/HCN ratio as a function of heliocentric distance than observed in 67P, so some contribution from an additional (nucleus) source seems likely. For NH$_2$CHO, the failure of all our models to reproduce the measured abundance in comet 67P by more than two orders of magnitude implies the presence of an additional source for this molecule --- for example, from the dissociation of NH$_4$COOH salt \citep{han19}, of from sublimation of NH$_2$CHO ice directly from the nucleus, albeit with a substantially lower abundance than found previously in Oort Cloud comets \citep{boc17}.

Our modeling work implies that HC$_3$N and NH$_2$CHO could be present in cometary nuclei at abundances that are significantly less than implied by previous (ground-based) observations. There is some evidence that the abundances of HC$_3$N and NH$_2$CHO are enhanced in typical cometary comae compared with the warm envelope of the solar-type protostar IRAS\,16293-2422, which is rich in the sublimated interstellar ices believed to be a major constituent of comets \citep{kah13,jab17,dro19}. On the other hand, the protostellar HC$_3$N/HCN ratio (which is easier to measure from the ground than the ratio with respect to H$_2$O) is only marginally smaller in IRAS\,16293-2422 (0.36\%; \citealt{dro19}) than in 67P (0.44\%; \citealt{han21}; full Rosetta mission). Gas-phase HC$_3$N/HCN ratios may be even larger in protoplanetary disks (3--134\%, depending on the assumed temperature; \citealt{ber18}), so it is likely that HC$_3$N can be incorporated into cometary nuclei during their accretion out of ice in the disk mid-plane. Comparison between cometary and interstellar/protostellar/disk abundances is nontrivial due to the possibility of chemical alteration during the passage of interstellar ices into the protoplanetary disk mid-plane and ultimately, into cometary nuclei. Indeed, HC$_3$N abundances are subject to significant modification by gas-phase processing in the warm regions surrounding the protostar \citep{wal14}. Nevertheless, our results show that chemical processing in the coma can lead to elevated abundances of some organic molecules, and this possibility should be taken into account when comparing abundances between interstellar ices and cometary comae, as a means for obtaining insights into the origin of cometary matter.  

Impacts between comets and planets were common during the early history of the Solar System, and may have allowed the delivery of organic molecules to planetary surfaces \citep[\emph{e.g.}][]{pie99,mcc14,tod20}. Both HC$_3$N and NH$_2$CHO have been implicated in the prebiotic synthesis of amino acids and DNA nucleobases on the primordial Earth \citep{fer68,sal12,pat15}. Our finding that these molecules may be less abundant in cometary ices than previously believed could therefore be of importance for theories concerning the chemical inventory of material available for the origin of life.

An important caveat concerning the extent of HC$_3$N and NH$_2$CHO synthesis in comets as a result of gas-phase coma chemistry is the uncertainty regarding the additional distributed CN sources ($P2$ and $P3$). The survey of \citet{del16} identified variability in the importance of the primary extended source of CN ($P2$) throughout the comet population, and the inner-coma CN source ($P3$) has only been detected in a single comet to-date (67P, although we are not aware of any dedicated attempts to detect a distinct, inner-coma CN source in any other comets). In our Model A, $P3$ was introduced as a photochemical source of CN, with a production rate and scale length tailored to reproduce the CN/HCN ratio measured by \citet{han20} in comet 67P. However, the detailed radial behaviour of this CN source is not yet well constrained, and it remains to be determined how well the CN abundance in Oort Cloud comets may be represented by the CN/HCN ratio in 67P.  Consequently, although we believe our models represent plausible scenarios for moderately-active OCCs, the specific, numerical results remain uncertain until inner-coma CN measurements are made in a larger sample of comets, and the properties of the dominant coma CN sources are fully elucidated.

\section{Conclusion}

Our chemical models for a moderately active comet, with organic-rich parent abundances, demonstrate that gas-phase synthesis of HC$_3$N and NH$_2$CHO occurs as a result of neutral-neutral reactions between simple precursor molecules known to be abundant in the coma. The presence of a distributed CN source in the inner coma, similar to that observed by Rosetta in comet 67P, increases the efficiency of these reactions, leading to effective HC$_3$N and NH$_2$CHO abundances in our model that can reach levels similar to those observed in moderately high-activity comets using ground-based millimetre-wave observations. Neutral-neutral coma chemistry can also reproduce the lower HC$_3$N/H$_2$O ratio measured by Rosetta in the relatively low-activity comet 67P before perihelion at $R_h\sim1.5$~au. As a result of such active gas-phase chemistry, the abundances of complex organics in the coma are therefore not necessarily representative of those stored in the nucleus ice. High-resolution mapping using millimetre/submillimetre facilities such as ALMA will be crucial for determining the importance of nucleus \emph{vs.} coma sources for these, and other complex organic molecules in future cometary apparitions. Additional observations, laboratory work, and modeling will be required to elucidate the distribution and origin of the inner-coma CN sources in comets other than 67P, and to determine the full impact of the dissociation products from ammonium salts and organic macromolecules on coma chemistry.

\section*{Acknowledgements}

This research was supported by the NASA Planetary Science Division Internal Scientist Funding Program through the Fundamental Laboratory Research work package (FLaRe).

\bibliography{Refs}{}
\bibliographystyle{mnras}

% Don't change these lines
\bsp	% typesetting comment
\label{lastpage}
\end{document}